# An improvement on fragmentation in Distribution Database Design Based on Knowledge-Oriented Clustering Techniques

VAN NGHIA LUONG
Department of Information Technology
Pham Van Dong University
Quang Ngai, Viet Nam
nghia.itq@gmail.com

HA HUY CUONG NGUYEN
Department of Information Technology
Quang Nam University
Quang Nam, Viet Nam
nguyenhahuycuong@gmail.com

VAN SON LE
Da Nang University of Education
Da Nang University
Da Nang, Viet Nam
levansupham2004@yahoo.com

*Abstract*—The problem of optimizing distributed database includes: fragmentation and positioning data. Several different approaches and algorithms have been proposed to solve this problem. In this paper, we propose an algorithm that builds the initial equivalence relation based on the distance threshold. This threshold is also based on knowledge- oriented clustering techniques for both of horizontal and vertical fragmentation. Similarity measures used in the algorithms are the measures developed from the classical measures. Experimental results carrying on the small data set match fragmented results based on the classical algorithm. Execution time and data fragmentation significantly reduced while the complexity of our algorithm in the general case is stable.

## I. INTRODUCTION

In distributed computing environments, each unit of data (item) which is accessed at the station, (site) is not usually a relationship but part of the relationship. Therefore, to optimize the performance of the query, the relations of global schema are fragmented into items.

There are several types of data fragmentation that are use vertical fragmentation, horizontal fragmentation, mixed fragmentation and derived fragments. Two classical algorithms associated with horizontal fragmentation and vertical fragmentation are PHORIZONTAL and BEA respectively [11].Many authors have proposed solutions improved the above two algorithms, as Navathe, et al., (1984) [13], Chakravarthy, et al., (1994) [3]..., However, the complexity of this algorithm is quite large, with vertical fragmentation problem is $O(n^2)$, where n is the number of attributes and horizontal fragmentation is $O(2^m)$, where m is the number of records [9],[11].

In recent years, several authors have incorporated to solve the problem of fragmentation and positioning, by using the optimal algorithms [5-6], [10] or using the heuristic method [4], [7]. The execution time of these algorithms is remarkably smaller than the classical algorithm.

The used technical association rules in data mining to vertical fragmentation has been mentioned in [8]. However, the data mining techniques do not attract many authors.

In this paper, we use knowledge-oriented clustering techniques for vertical and horizontal fragmentation problem. The measure of similarity was developed based on from the available measure of the classical algorithms in data mining.

In the clustering algorithm based on knowledge-oriented, we propose an algorithm that builds the initial equivalent relation based on the distance threshold. This approach differs from the previous works proposed by Hirano et al.[10], [14] and Bean et al.,[2], [14-16] in that the proposed algorithm automatically determines the number of clusters based on the data set of survey.

The paper is organized as follows: section 2 presents a brief overview of the basic concepts. We detail with the proposed vertical and horizontal fragmentation algorithms, in section 3 and section 4, respectively. We then discuss the main contributions of proposed approach in section 5.

## II. BASIC CONCEPTS

### A. Vertical fragmentation

Vertical fragmentation is the collective decay properties of the relational schema R into the sub schema $R_1, R_2, ..., R_m$, such that each attribute in these sub schemas is often accessed together.

To show how often the same queries together, Hoffer and Severance introduced the concept attribute affinity [11].

If $Q = \{q_1, q_2, .., q_m\}$ is a set of applications, $R(A_1, A_2, ..., A_n)$ is a relational schemas. The relationship between $q_i$ and attributes $A_j$ is determined by using the values:

$$use(q_i, A_j) = \begin{cases} 1, & A_j \text{ is engaged in } q_i \\ 0, & A_j \text{ is not engaged in } q_i \end{cases} \quad (1)$$

Put $(A_i, A_j) = \{q \in Q \mid use(q, A_i) \cdot use(q, A_j) = 1\}$. Attribute affinity between $A_i$ and $A_j$ is:

$$Aff(A_i, A_j) = \sum_{q \in Q(A_i, A_j)} \sum_{\forall S_l} (\sum ref_l(q) * acc_l(q)) \quad (2)$$

In particular, refl(q): the number of pairs of attributes $(A_i, A_j)$ is referenced in the application q at station $S_l$; accl(q): frequency of access to applications q in station $S_l$. BEA algorithm consists of two main phases:

*1) Permutations row, column affinity matrix of attribute to obtain the cluster affinity matrix (CA) which has global affinity measure AM (global affinity measure) [11] is the largest.*

*2) Find the partition of the set of attributes from the matrix CA by exhaustive method, so that:*
Z= CTQ *CBQ – COQ2 is the maxima, with:

$$CTQ = \sum_{q \in TQ} \sum_{\forall S_j} ref_j(q_j) acc_j(q_i)$$

$$CTQ = \sum_{q \in TQ} \sum_{\forall S_j} ref_j(q_j) acc_j(q_i)$$

$$COQ = \sum_{q \in OQ} \sum_{\forall S_j} ref_j(q_j) acc_j(q_i)$$

TABLE I.  CLUSTER AFFINITY MATRIX CA

|      | $A_1$ | $A_2$ | .. | $A_i$ | $A_{i+1}$ | .. | $A_n$ |
|------|-------|-------|-----|-------|-----------|-----|-------|
| $A_1$ |       |       |     |       |           |     |       |
| ..   |       | TA    |     |       |           |     |       |
| $A_i$ |       |       |     |       |           |     |       |
| $A_{i+1}$ |   |       |     |       |           |     |       |
|      |       |       |     |       | BA        |     |       |
| $A_n$ |      |       |     |       |           |     |       |

In which,

AQ($q_i$)= {$A_j$| use($q_i$, $A_i$)=1};

TQ={$q_i$ | AQ($q_i$) ⊆TA};

BQ= {$q_i$ | AQ($q_i$) ⊆BA};

OQ=Q\ {TQ ∪BQ}

The complexity of the algorithm is proportional to $n^2$.

### B. Horizontal Fragmentation

Horizontal fragmentation divides set records into a smaller set of records. Horizontal fragmentation is based on the query conditions, which are expressed through simple predicates of the form: $A_j \theta$<value>.

Set Pr = {$Pr_1$, $Pr_2$, ..,$Pr_k$} is a set of simple predicates extracted from a set of applications. A conjunction of the predicates, which is built from $P_r$ will have the form:

$$P_1^* \wedge P_2^* \wedge ... \wedge P_n^* \quad (3)$$

Where $P_i^*$ is a predicate, which received one of $P_i$ or $\neg P_i$ values.

PHORIZONTAL algorithm uses the conjunction of the predicates $P_1^* \wedge P_2^* \wedge ... \wedge P_n^*$ to find the conditions for horizontal fragmentation of data [9]. The relation r(R) will be fragmented into {$r_1(R), r_2(R),..,r_k(R)$}, with $r_i(R) = \sigma\ F_i(r(R))$, $1 \leqslant i \leqslant k$; $F_i$ is a predicate, which forms the conjunction of the primary predicates [9].

### C. Information systems and the inability to distinguish relationship

- The information system is a pair of SI = (U, A), where U is a finite set of objects U={$t_1, t_2, ..,t_n$}, A is non-empty finite set of attributes.

- An equivalent relation (A binary relations satisfy properties reflective, symmetric and transitive) defined on U is called an inability to distinguish relationship (irrespectively relationship) on U.

### D. Clustering algorithm Knowledge-Oriented

Clustering algorithm Knowledge-Oriented based on rough set theory was first proposed by the authors Shoji Hirano, et al., [10], [15]. This is a clustering algorithm automatically determines the number of clusters based on the survey data [12]. The main idea of this clustering algorithm consists of 2 phases:

*1) Created of equivalence relation on the set of object clustering.*

*2) Editing of the equivalence relation using a threshold Tk based on the measure irrespective. This iterative process will update Tk a best clustering results is obtained Using this algorithm to data fragmentation, we have proposed initial equivalence relation based on the average distance between objects.*

Clustering algorithm Based on knowledge orientation, so we propose as follows:

**Input**: U= the set of objects to be clustered.

(Each object must be describe the information needed to construct a similar measure).

**Output**: The clusters (corresponding to the fragment of data).

**Method**:

*Step 1*: Construct a matrix of similarities $S=S(t_i, t_j)$ between all pairs of objects $(t_i, t_j)$;

*Step 2*: Specify a initial ability to distinguish relationship $R_i$ for each object. Synthesis to get an initial clustering;

*Step 3*: Construct ability to distinguish matrix $\Gamma=\gamma(t_i, t_j)$ to assess the quality of clustering;

*Step 4*: Modify the clusters by the inability to distinguish relationship $R_j^{mod}$ for each object to achieve the revised clustering;

*Step 5*: Repeat steps 3 and 4 until a stable clustering is obtained.

The inability to distinguish relationship corresponds to the $i^{th}$ attribute:

$R_i = \{(t_i, t_j) \in U \times U : d(t_i, t_j) \leq Tk_j, \text{ with } j = 1, 2, \ldots, n\}$.

Where $d(t_i, t_j)$ is the distance between two participants clustering.

The threshold $Tk_j$ is determined as follows:

$$Tk_i \left[ \sum_{j=1, j \neq i}^{n} (1 - s(t_i, t_j)) \right] / (n-1) \quad (4)$$

With $s(t_i, t_j)$ is the similarity measure of two objects $t_i, t_j$.

### III. VERTICAL FRAGMENTATION PROBLEM BASED ON KNOWLEDGE-ORIENTED CLUSTERING TECHNIQUES

Vertical fragmentation problem is converted to the clustering problem, based on the following concepts:

#### A. Attribute and the reference feature vector

**Definition 1:** The reference measure of transaction $q_i$ on attribute $A_j$, denoted by $M(q_i, A_j)$:

$$M_{ij} = M(q_i, A_j) = use(q_i, A_j) * f_i$$

In which $M_{ij}$ is the frequency with which transactions $q_i$ reference to attribute $A_j$. With $f_i$ is the frequency of transactions $q_i$ and $use(q_i, A_j)$ is defined by formula (1).

**Definition 2:** $VA_j$ reference feature vector of attribute $A_j$ with reference transactions $(q_1, q_2, \ldots, q_m)$ is defined as follows:

|  | $q_1$ | $q_2$ | … | $q_m$ |
|---|---|---|---|---|
| $VA_j =$ | $M_{1j}$ | $M_{2j}$ | … | $M_{mj}$ |

#### B. The similarity measure of two properties

**Definition 3**: The similarity measure of two attributes $A_k$, $A_l$ has two feature vectors corresponding to the reference transactions $(q_1, q_2, \ldots, q_m)$:

$VA_k = (M_{1k}, M_{2k}, \ldots, M_{mk})$

$VA_l = (M_{1l}, M_{2l}, \ldots, M_{ml})$

Is determined by the cosine measure:

$$s(A_k, A_l) \frac{VA_k * VA_l}{\|VA_k\| * \|VA_l\|} = \frac{\sum_{i=1}^{m} M_{ik} * M_{il}}{\sqrt{\sum_{i=1}^{m} M_{ik}^2} * \sqrt{\sum_{i=1}^{m} M_{il}^2}} \quad (5)$$

#### C. Vertical fragmentation based on knowledge-oriented clustering technique

To illustrate the vertical fragmentation algorithm based on knowledge-oriented clustering techniques. We use the assumption of examples about vertical fragmentation problem based on BEA algorithm is presented in [1], [8]:

The set of attributes $A_t = \{A_1, A_2, A_3, A_4\}$

The set of transactions $Q = \{q_1, q_2, q_3, q_4\}$. The matrix used:

|  | $A_1$ | $A_2$ | $A_3$ | $A_4$ |
|---|---|---|---|---|
| $q_1$ | 1 | 0 | 1 | 0 |
| $q_2$ | 0 | 1 | 1 | 0 |
| $q_3$ | 0 | 1 | 0 | 1 |
| $q_4$ | 0 | 0 | 1 | 1 |

The frequency of application execution with a set of transactions $\{q_1, q_2, q_3, q_4\}$, and $F = \{f_1, f_2, f_3, f_4\} = \{45, 5, 75, 3\}$.

From the assumption, we have the reference feature vectors:

|  | $q_1$ | $q_2$ | $q_3$ | $q_4$ |
|---|---|---|---|---|
| $VA_1 =$ | 45 | 0 | 0 | 0 |
| $VA_2 =$ | 0 | 5 | 75 | 0 |
| $VA_3 =$ | 45 | 5 | 0 | 3 |
| $VA_4 =$ | 0 | 0 | 75 | 3 |

The similar matrix $S_{4 \times 4} = (s(A_k, A_l))$ $k = 1,4; l = 1,4$

|  | $A_1$ | $A_2$ | $A_3$ | $A_4$ |
|---|---|---|---|---|
| $A_1$ | 1 | 0 | 0.9918 | 0 |
| $A_2$ |  | 1 | 0.0073 | 0.9970 |
| $A_3$ |  |  | 1 | 0.0026 |
| $A_4$ |  |  |  | 1 |

The result of the vertical fragmentation algorithm based on the clustering algorithm towards knowledge-oriented.

| Cluster | Set of attributes |
|---|---|
| 1 | $\{A_1, A_3\}$ |
| 2 | $\{A_2, A_4\}$ |

This fragmentation results correlate with the results of the vertical fragmentation by algorithm BEA.

## IV. HORIZONTAL FRAGMENTATION PROBLEM BASED ON KNOWLEDGE-ORIENTED CLUSTERING TECHNIQUE

Similar to vertical fragmentation, assuming conversion of horizontal fragmentation problem from PHORIZONTAL algorithm is based on the concept of the following establishments:

### A. Vectorization records of a relationship

Considering the relations $r(R) = \{T_1, T_2, .., T_l\}$, the set of simple predicates extracted from applications on $r(R)$ is $Pr = \{Pr_1, Pr_2, .., Pr_m\}$. Vector binary of records by the rule:

TABLE II. VECTORIZATION BINARY

|  | $P_{r1}$ | $P_{r2}$ | .. | $P_{rj}$ | .. | $P_{rm}$ |
|---|---|---|---|---|---|---|
| $T_1$ | $a_{11}$ | $a_{12}$ | .. | $A_{1j}$ | .. | $a_{1m}$ |
| .. |  |  | .. |  |  | .. |
| $T_i$ | $a_{i1}$ | $a_{i2}$ | .. | $a_{ij}$ | .. | $a_{im}$ |
| .. |  |  | .. |  |  | .. |
| $T_l$ | $a_{l1}$ | $a_{l2}$ | .. | $a_{lj}$ | .. | $a_{lm}$ |

$$\forall a_{ij} = \begin{cases} 1, & \text{if } T_i[Pr_j] = true \\ 0, & \text{if } T_i[Pr_j] = false \end{cases}$$

### B. The similarity measure of two binary vector

Consider two vectors $x_i$ and $x_j$, that are represented by binary variables. Assuming binary variables have the same weight. We have event tables as Table 3. Where q is the number of binary variables equal to 1 for the two vectors $x_i$ and $x_j$, s is the number of binary variables equal to 0 fo simpler $x_i$ but equal to 1 for $x_j$, r is the number of binary variables equal to 1 for $x_i$ but is 0 for $x_j$, t is the number of binary variables equal to 0 for all vectors $x_i$ and $x_j$.

TABLE III. EVENT TABLE FOR BINARY VARIABLES

|  |  | Object j |  |  |
|---|---|---|---|---|
|  |  | *1* | *0* | *Sum* |
| Object i | *1* | q | r | q+r |
|  | *0* | s | t | s+t |
|  | *Sum* | q+s | r+t | p |

- The difference of two vectors $x_i$ and $x_j$ based on the symmetric binary dissimilarity are:

$$d(x_i, x_j) = \frac{r+s}{q+r+s+t} \quad (6)$$

- The similarity measure between two vectors $x_i$ and $x_j$ is defined by the Jaccard coefficient:

$$sim(x_i, x_j) = 1 - d(x_i, x_j) \quad (7)$$

### C. Horizontal fragmentation based on knowledge-oriented clustering techniques

Considering the relations EMP, [11]:

TABLE IV. SAMPLE DATA (EMP) FOR HORIZONTAL FRAGMENTATION

|  | ENo | EName | Title |
|---|---|---|---|
| $T_1$ | $E_1$ | Jjoe | Elect-Eng |
| $T_2$ | $E_2$ | M.Smith | Syst-Analyst |
| $T_3$ | $E_3$ | A.Lee | Mech-Eng |
| $T_4$ | $E_4$ | J.Smith | Programmer |
| $T_5$ | $E_5$ | B.Casey | Syst-Analyst |
| $T_6$ | $E_6$ | L.Chu | Elect-Eng |
| $T_7$ | $E_7$ | R.David | Mech-Eng |
| $T_8$ | $E_8$ | J.Jone | Syst-Analyst |

Consider two simple predicates:

$p_1 = $(Title > "Programmer"); $p_2 = $(Title < "Programmer"), with string comparison rules in alphabetical order. Vectorization records by two predicates $p_1$ and $p_2$ are:

TABLE V. VECTORIZATION RECORDS

|  | $p_1$ | $p_2$ |
|---|---|---|
| $T_1$ | 1 | 0 |
| $T_2$ | 0 | 1 |
| $T_3$ | 1 | 0 |
| $T_4$ | 0 | 0 |
| $T_5$ | 0 | 1 |
| $T_6$ | 1 | 0 |
| $T_7$ | 1 | 0 |
| $T_8$ | 0 | 1 |

### D. The result horizontal fragmentation of the relation (EMP)

The result horizontal fragmentation of the relation EMP as Table III based on the clustering algorithm towards knowledge-oriented. We have used the similarity measure defined by formula (7), where $d(x_i, x_j)$ is calculated by the formula (6).

With k=2, we have:

| Cluster | The set of records |
|---|---|
| 1 | $T_1, T_3, T_6, T_7$ |
| 2 | $T_2, T_4, T_5, T_8$ |

With k=3, we have:

| Cluster | The set of records |
|---|---|
| 1 | $T_1, T_3, T_6, T_7$ |
| 2 | $T_2, T_5, T_8$ |
| 3 | $T_4$ |

And with k=4, we have:

| Cluster | The set of records |
|---------|--------------------|
| 1 | $T_1, T_3$ |
| 2 | $T_2, T_5, T_8$ |
| 3 | $T_4$ |
| 4 | $T_6, T_7$ |

This fragmentation results coincide with the results of the horizontal fragmentation by algorithm PHORIZONTAL.

## V. CONCLUSION

In this paper, we used knowledge-oriented clustering techniques for fragmentation problem in distributed database systems. With this solution, we also proposed transforming hypothetical of this problem to hypothetical of clustering problems. Experimental results on the data in [11] that results correlate with the results obtained from the classical fragmentation algorithms PHORIZONTAL and BEA. In addition to experimental data as presented, we also tested on a number of different data set. The results are also similar to the two classical algorithms above. In the future work, we will carry out the Analysis of large data sets that to compare test the usability of the proposed solution.

**Personal information:**

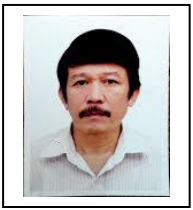

  - Full name: **Luong Van Nghia**, year of birth 1964.

  - Address: 99 Hung Vuong street, Quang Ngai city, Quang Ngai province, Vietnam.

  - Academic degree: Master of Computer Science

  - Name of Agency: Faculty of Informatics, Pham Van Dong Universisty

  - Phone:: (+84) 913 498 804, email: nghia.itq@gmail.com  (email2: lvnghia@pdu.edu.vn)

  - Graduation year: Master in Computer Science, Hue University of Science

  - Research Areas: Data mining, Distributed Database, Embedded System.